\documentstyle[12pt,aaspp4]{article}

\begin{document}

\title{A NEW DUAL-COMPONENT PHOTOIONIZATION MODEL
       FOR THE NARROW EMISSION-LINE REGIONS
       IN ACTIVE GALACTIC NUCLEI}
   
\bigskip
\author{\sc Takashi Murayama \& Yoshiaki Taniguchi}

\bigskip
\affil{Electronic mail: murayama@astr.tohoku.ac.jp,
       tani@astr.tohoku.ac.jp}

\bigskip

\affil{Astronomical Institute,
        Tohoku University,
         Sendai 980-8578, Japan}

\authoremail{murayama@astr.tohoku.ac.jp}


\begin{abstract}
Having found that type 1 Seyfert nuclei have excess
[\ion{Fe}{7}] $\lambda$6087 emission with respect to type 2s,
Murayama \& Taniguchi have proposed that the high-ionization
nuclear emission-line region (HINER) traced by the
[\ion{Fe}{7}] $\lambda$6087 emission resides in the inner wall of
dusty tori.
The covering factor of the torus is usually large (e.g., $\sim 0.9$).
Further, electron density in the tori (e.g., $\sim 10^{7\mbox{--}8}$ cm$^{-3}$)
is considered to be higher significantly than that
(e.g., $\sim 10^{3\mbox{--}4}$ cm$^{-3}$) in the narrow-line region (NLR).
Therefore it is expected that the torus emission contributes to the
majority of the higher-ionization emission lines. Taking this HINER
component into account, we have constructed new dual-component (i.e.,
a typical NLR with a HINER torus) photoionization models.
Comparison of  our model  with the observations show that  that if the
torus emission contributes  $\sim 10$ \% of the NLR emission,
our dual-component model  can explain the observed high
[\ion{Fe}{7}] $\lambda$6087/[\ion{O}{3}] $\lambda$5007 intensity
ratios of the Seyfert 1s without invoking any unusual assumptions
(e.g., the overabundance of iron).
\end{abstract}



\section{INTRODUCTION}
It has been often considered that
emission-line regions around active galactic nuclei (AGN) are
photoionized by the non-thermal continuum radiation from central engines
(Davidson 1977; Kwan \& Krolik 1981; see for a review 
Davidson \& Netzer 1979; Osterbrock 1989).
However, this photoionization scenario has been sometimes confronted with
several serious problems; 1) any single-cloud photoionization models
underpredict
the [\ion{O}{3}] $\lambda$4363/[\ion{O}{3}] $\lambda$5007 intensity ratio
(e.g., Filippenko \& Halpern 1984),
2) any (ionization-bounded) single-cloud models cannot explain the large
temperature difference between $T_{\rm [OIII]}$ and
$T_{\rm [NII]}$
(e.g., Wilson, Binette, \& Storchi-Bergmann 1997),
3) any single-cloud photoionization models
underpredict higher ionization emission lines including ultraviolet 
emission lines such as \ion{C}{3}] $\lambda$1909 and
\ion{C}{4} $\lambda$1549
(e.g., Pelat, Alloin, \& Bica 1987; see also Dopita et al.\ 1997),
and 4) any single-cloud photoionization models require the overabundance
in some particular elements such as nitrogen and iron (e.g., Ho,
Shields, \& Filippenko 1993; Hamann \& Ferland 1992, 1993). 
In order to reconcile the above dilemmas, multi-component cloud models have 
been proposed (Stasi\'nska 1984a, 1984b; Filippenko \& Halpern 1984;
Binette 1985; Ferland \& Osterbrock 1986;
Binette, Robinson, \& Couvoisier 1988; Binette et al.\ 1996;
Komossa \& Schulz 1997;
Ferguson, Korista, \& Ferland 1997a; Ferguson et al.\ 1997b).
Although this approach 
seems to be the natural extension of the single-cloud photoionization models,
there is no guiding principle for the multi-component modeling
because we do not have any unambiguous observational constraint
on the geometrical structure of the ionized clouds in AGN.

Recently Murayama \& Taniguchi (1998, hereafter MT98) have found that
type 1 Seyfert nuclei (S1s) have excess [\ion{Fe}{7}] $\lambda$6087
emission with respect to type 2s (S2s). The S1s exhibit broad emission 
lines which are attributed to ionized gas very close to the central
engine (e.g., $\sim 0.01$ pc),
whereas the S2s do not show such broad lines. The current
unified model of AGN explains this difference as that the broad line
region in the S2s is hidden from the line of sight by a dusty torus if 
we observe it from a nearly edge-on view toward the torus.
Therefore, the finding of MT98 implies that the high-ionization
nuclear emission-line region (HINER: Binette 1985;
Murayama, Taniguchi, \& Iwasawa 1998) traced by
the [\ion{Fe}{7}] $\lambda$6087 emission resides in the inner wall of
such dusty tori. Since the covering factor of the torus is usually
large (e.g., $\sim 0.9$), and the electron density in the tori
(e.g., $\sim 10^{7\mbox{--}8}$ cm$^{-3}$)
is considered to be significantly higher than that
(e.g., $\sim 10^{3\mbox{--}4}$ cm$^{-3}$) of the narrow-line region (NLR),
the contribution from the torus dominates the emission of the
higher-ionization lines (Pier \& Voit 1995).
Taking this HINER component into account, we  construct new
dual-component (i.e., a typical NLR with a HINER torus) photoionization
models and examine whether  these models can explain the observations without
invoking any unusual assumptions.

\section{PHOTOIONIZATION MODELS}
\subsection{Single Cloud Models}

In order to investigate the HINER  of Seyfert
galaxies within the framework of photoionization, 
we first perform single-cloud photoionization model
calculations using the spectral synthesis 
code CLOUDY version 90.04 (Ferland 1996)
which  solves the equations of statistical and
thermal equilibrium and produces a self-consistent
model of the run of temperature as a function of depth into the nebula.
Here we assume that a uniform density, dust-free gas
cloud with plane-parallel geometry is ionized by 
a power-law continuum source.

The parameters for the calculations are 1) the hydrogen density
of the cloud ($n_{\rm H}$),
2) the ionization parameter ($U$), 3) the spectral index of
ionizing continuum between 10 $\mu$m and 250 \AA~ ($\alpha$; $f_\nu 
\propto \nu^{\alpha}$),
and 4) the chemical compositions.
We adopt the following continuum spectrum taking account of
the actually observed one in AGNs (cf.\ Ho, Shields,
\& Filippenko 1993); 
1) $\alpha = 2.5$ for $\lambda > 10 \mu$m,
2) $\alpha = -2$ -- $-1$ for 250 \AA $ < \lambda \leq 10$ $\mu$m, and  
3) $\alpha = -2$ for $\lambda \leq 250$ \AA. 
The total amount of metals is expressed as $Z$ in units of $Z_\odot$.
In this Letter, the abundance of each element is defined as a number fraction relative to
hydrogen and is expressed by bracketed quantities ;
e.g., ${\rm [Fe]}=n({\rm Fe})/n({\rm H})$.
All the abundances are in units of solar values.
However, taking into account the many lines of evidence in favor of a
nitrogen overabundance  (cf.\ Storchi-Bergman et al.\ 1996;
Hamann \& Ferland 1992, 1993), we adopt twice  the solar nitrogen abundance
as the standard value in our model calculations following 
Ho, Shields, \& Filippenko (1993).
Namely, all  elements  have solar values
except for  nitrogen
whose abundance will be twice  the solar value.
Adopted solar abundances relative to hydrogen are taken from
Grevesse \& Anders (1989).
The calculations were  stopped when the temperature fell to 4000 K,
below which it was verified that little optical emission took place.

We are not going to present detailed description of these single-cloud 
models because the main purpose of this Letter is to present new
dual-component photoionization models.
Therefore, we give a summary of the single-cloud models briefly (see
for details Murayama 1998) and describe the difficulty of the models.
We have performed several model runs covering the following parameter ranges:
a) $\alpha=-2$, $-1.5$, and $-1$, b) $\log U = -2.5$, $-2$, $-1.5$,
and $-1$, c) $\log n_{\rm H}$ (cm$^{-3}$) $= 4$, 5, 6, and 7, d) $[Z]=1$, 2, 5, and
10, and e) $[{\rm Fe}]=1$, 2, 5, and 10 keeping the other elemental
abundances to the  solar values.
These parameter ranges were adopted in order to reproduce the observational
properties of the HINER. We have confirmed that the intensity of
[\ion{Ca}{5}] $\lambda$6087 is always less than 5\% of that of
[\ion{Fe}{7}] $\lambda$6087 in all the cases, providing the validity
of the use of [\ion{Fe}{7}] $\lambda$6087 emission line.

Using several newly defined diagnostic diagrams
(e.g., [\ion{Fe}{7}] $\lambda$6087/[\ion{O}{3}] $\lambda$5007
vs.\ [\ion{O}{1}] $\lambda$6300/[\ion{O}{3}] $\lambda$5007,
[\ion{Fe}{10}] $\lambda$6374/[\ion{Ne}{5}] $\lambda$3426
vs.\ [\ion{Fe}{7}] $\lambda$6087/[\ion{Ne}{5}] $\lambda$3426, and
[\ion{Fe}{10}] $\lambda$6374/[\ion{O}{3}] $\lambda$5007
vs.\ [\ion{Fe}{7}] $\lambda$6087/[\ion{O}{3}] $\lambda$5007),
we have compared the model results with the observations (Osterbrock
1977, 1985; Koski 1978; Osterbrock \& Pogge 1985; Shuder \& Osterbrock 
1981; Murayama 1998) and found that; 1) the most probable values of
$\log U$ and $\alpha$ are $-2$ and $-1$, respectively,
2) $n_{\rm H} \sim 10^{4\mbox{--}7}$ cm$^{-3}$,  
and 3) an unusually high iron abundance, up to ten times  the
solar value while keeping the remaining elements the solar values,
was required to explain  the observed very high
[\ion{Fe}{7}] $\lambda$6087/[\ion{O}{3}] $\lambda$5007
ratios for the S1s (Murayama 1998).
To briefly demonstrate these results, a comparison between the
observations and the models 
is shown in Figure 1 which is a  diagram of 
[\ion{Fe}{7}] $\lambda$6087/[\ion{O}{3}] $\lambda$5007
vs.\ [\ion{O}{1}] $\lambda$6300/[\ion{O}{3}] $\lambda$5007.
Note that we adopt $\log U = -2$ and $\alpha = -1$ for all the models
presented here. A serious problem is that a wide span in the iron
abundance extending up to ten times  solar is required
to explain the observational data points. 
Although the circumnuclear star-forming regions in Seyfert nuclei show 
a nitrogen overabundance of up to a factor of three at most (e.g.,
Storchi-Bergmann et al.\ 1996), it is quite unlikely that the NLR in
the Seyfert nuclei shows a factor of ten  overabundance of iron
because this overabundance cannot be explained in the terms of
ordinary star formation history in galaxies.
We also mention that
single cloud models predict
a tendency towards smaller [\ion{O}{1}]/[\ion{O}{3}]
with increasing iron abundance.
This may be due to the fact  that when one increases the iron abundance,
the contribution of Fe$^{0}$ and Fe$^{+}$ to 
the cooling within the partly ionized region increases more
than that of O$^{0}$.

\subsection{Dual-Component Cloud Models}
As shown in MT98,
the observed [\ion{Fe}{7}]/[\ion{O}{3}] ratios are systematically
higher in  S1s than in  S2s.
This suggests that the major contributor to the HINER emission
in  Seyferts has a viewing angle dependence in accordance with current unified models
of AGN (e.g., Antonucci 1993). It is reasonable to consider that 
the most probable site of this HINER emission corresponds to the inner edges of dusty tori.
We therefore proceed to construct dual-component models
in which the inner surface of a torus is introduced as
a new ionized-gas  component in addition to the traditional
NLR  component.
The observed [\ion{Fe}{7}]/[\ion{O}{3}]
ratios of the S1s are $\sim 0.1$ on the average
while those of the S2s are $\sim 0.01$ (MT98). 
It is hence suggested that the typical NLRs have intensity
ratios of [\ion{Fe}{7}]/[\ion{O}{3}] $\sim 0.01$ and that
the observed higher ratios in S1s than in S2s are mostly due to the
contribution from the HINER torus.

For the HINER torus component, we calculate photoionization models
as follows.
The single-cloud model suggests that the ionization parameter lies in
the range of $\log U \simeq -1.5$ -- $-2$.
As for the electron density, it is often considered that the inner
edges of tori have higher electron densities, e.g.,
$n_{\rm e} \sim 10^{7\mbox{--}8}$ cm$^{-3}$ (Pier \& Voit 1995).
This higher density is also expected in terms of the locally
optimally-emitting cloud (LOC) models (Ferguson et al.\ 1997a).
We therefore calculated models covering the
following  ranges of parameters; $\log U = -2.2$, $-2.0$, $-1.8$, $-1.6$,
and $-1.4$, and $\log n_{\rm H}$ (cm$^{-3}$) = 6, 6.5, 7, 7.5, 8, 8.5, and 9.
In order to increase the 
[\ion{Fe}{7}]/[\ion{O}{3}] ratio by one order of magnitude, the
contribution  to the [\ion{Fe}{7}] emission of the torus component
must be very high.
Because the largest [\ion{Fe}{7}]/[\ion{O}{3}]
ratio of the observed data is $\sim 0.5$, [\ion{Fe}{7}]/[\ion{O}{3}]
of the torus component must be greater than 0.5.
However, we find that ionization-bounded models 
cannot explain the observed large [\ion{Fe}{7}]/[\ion{O}{3}] values
by simply increasing electron densities up to $10^{9}$ cm$^{-3}$.
Further, such very high-density models yield unusually strong [\ion{O}{1}]
emission with respect to [\ion{O}{3}].
We therefore assume ``truncated'' clouds with both large
[\ion{Fe}{7}]/[\ion{O}{3}]
ratios and little low-ionization lines for the HINER torus.
The calculations were stopped at a hydrogen column density
when [\ion{Fe}{7}]/[\ion{O}{3}] $=1$.
The results are summarized in Table 1.
For all the models, we also calculated
an [\ion{Fe}{10}] $\lambda$6374/[\ion{Fe}{7}] $\lambda$6087 ratio.
Since the average 
[\ion{Fe}{10}]/[\ion{Fe}{7}] ratio for Seyferts is
$\sim$ 0.4 (e.g., Pier \& Voit 1995),
the models with  [\ion{Fe}{10}]/[\ion{Fe}{7}] $\gg 1$
are not appropriate. 
Also, taking into account  the critical density of [\ion{Fe}{7}],
3.6$\times 10^7$ cm$^{-3}$ (De Robertis \& Osterbrock 1984),
we adopt the model with $n_{\rm H} = 10^{7.5}$ cm$^{-3}$ and $\log U = -2.0$,
which gives a ratio [\ion{Fe}{10}]/[\ion{Fe}{7}] of 0.8 as a
representative model for the HINER torus.
This model has a characteristic thickness of 
$l = N_{\rm H}/n_{\rm H} \simeq 10^{12.9}$ cm.
Therefore, if we regard that the cloud is matter bounded,
its size is estimated to be $\sim 10^{13}$ cm.
On the other hand, if we regard the cloud as an ionization bounded one,
we have to consider a case that the cloud is so dense and dusty
that only its surface is photoionized.
This is indeed the cloud model considered by Pier \& Void (1995).
Their model gives a typical thickness of the surface layer at
UV dust optical depth $\tau_{\rm d} \sim 1$,
$l_{\rm layer}\sim 6 \times 10^{12}$ $F^{-1}_7$ $T_4$
$\chi^{-1}_{\rm d}$ cm, where $F_7$ is the
ionizing energy flux in units of $10^7$ erg cm$^{-2}$ s$^{-1}$,
$T_4$ is the gas temperature in units of $10^4$ K, and 
$\chi_{\rm d}$ is the dust-to-gas ratio relative to the Galactic
value.
This is nearly the same as the characteristic length of our best model.
Therefore our truncated models can be consistently understood in
terms either of matter bounded models or of ionization bound models.

Now we can construct dual-component models combining this torus
component model with the NLR models with $\alpha=-1$, $\log U=-2$,
$[{\rm Fe}]=1$, and $\log n_{\rm H}$ (cm$^{-3}$) = 0 -- 7 (see
section 2.1). 
When we construct a composite model between the NLR and the
torus component, we have to take into account  the covering factor of the 
torus. It is known that the NLR shows a biconical emission-line
morphology with a  typical semi-opening angle of about 30\arcdeg{}
(e.g., Pogge 1989; Wilson \& Tsvetanov 1994; Schmitt \& Kinney
1996). This semi-opening angle implies a covering factor of the torus,
$1-\Delta\Omega/(4\pi)\simeq 0.9$ where $\Delta\Omega$ is the full
opening solid angle of the NLR cones in 
steradian. Therefore, one can infer that the relative contribution of the NLR component
may be of the order  10 \%. In Figure 2, we present the results of the
dual-component models. Here the lowest dashed line shows the results of the
NLR component models with $\alpha=-1$, $\log U=-2$,
and $[{\rm Fe}]=1$ as a function of $n_{\rm H}$ from
1 cm$^{-3}$ to $10^{6}$ cm$^{-3}$. If we allow  the contribution
from the torus component to reach up to $\sim 50$ \% in the Seyferts
with very high  [\ion{Fe}{7}]/[\ion{O}{3}] ratios,
we can explain all the data points without invoking the unusual iron
overabundance.
Note that the majority of objects can be explained by simply
introducing a $\sim 10$ \%
contribution from the HINER torus.
The important quantities characterizing the models are summarized in 
Table 2.
The model area shown in Figure 2 fits to the observed data points
better than that in Figure 1.
The dual-component models predict the occurrence of a lower cut-off 
in the [\ion{O}{1}]/[\ion{O}{3}] ratio,
i.e., no object with $\log$ [\ion{O}{1}]/[\ion{O}{3}] smaller than
$-1.8$, which is consistent with the observations.
However, as shown in Figure 1,  single cloud models
of both low density and high iron abundance
predict on the contrary the existence of AGNs with 
$\log$ [\ion{O}{1}]/[\ion{O}{3}] $<-1.8$.
This further supports our contention that the dual-component models studied here are
more successful than the single cloud models.

\section{CONCLUDING REMARKS}

There has been a debate about the origin of the high-ionization lines.
The possible origins are summarized as follows:
1) a hot collisionally ionized gas with temperatures of $T_{\rm e} \sim 10^6$ K
(Oke \& Sargent 1968; Nussbaumer \& Osterbrock 1970),
2) a cool gas ($T_{\rm e} \sim$ a few $\times 10^4$ K and $n_{\rm H}
\sim 10^6$ cm$^{-3}$)
photoionized by the central non-thermal continuum emission (Osterbrock 1969;
Nussbaumer \& Osterbrock 1970; Grandi 1978),
3) a low-density  (e.g., $n_{\rm H} \sim 1$ cm$^{-3}$) interstellar
medium (ISM) photoionized by the central non-thermal continuum
emission (Korista \& Ferland 1989;
see also Murayama, Taniguchi, \& Iwasawa 1998),
and 4) a combination of shocks and photoionization by the central non-thermal
continuum emission (Viegas-Aldrovandi \& Contini 1989).
Although  photoionization models seem to be a modest idea,
they cannot explain the observed strong high-ionization lines
(Murayama, Taniguchi, \& Iwasawa 1998 and references therein).
However, MT98 found that the S1s have excess [\ion{Fe}{7}] $\lambda$6087
emission with respect
to the S2s, implying that the majority of the HINER emission
in the S1s arises from
the inner wall of dusty tori. Indeed we have shown that our new dual-component
photoionization models incorporating the contribution from the HINER torus
can explain the observed higher [\ion{Fe}{7}]/[\ion{O}{3}] ratios in the S1s.
Although shock models can also explain the observations 
(Dopita \& Sutherland 1995, 1996; Dopita et al.\ 1997; Terlevich et al.\ 1992),
it is important to construct more realistic multi-component
photoionization models
in which detailed physical properties of ionized clouds are taken into account
(e.g., ionization-bounded and matter-bounded clouds:
Binette et al.\ 1996, 1997; Wilson et al.\ 1997).

\acknowledgements
We would like to thank Gary Ferland for providing us his
CLOUDY code and B.\ Vila-Vilalo for useful discussions.
We would also like to thank the referee, Luc Binette
for useful comments and suggestion.
TM was supported by the Grant-in-Aid for JSPS Fellows
by the Ministry of Education, Science, Sports and Culture.
This work was financially supported in part by Grant-in-Aids for the Scientific
Research (No. 07044054) of the Japanese Ministry of
Education, Science, Sports and Culture. 
\clearpage


\clearpage


\begin{deluxetable}{ccccc}
\tablecaption{A summary of the torus models}
\tablehead{
\colhead{$\log n_{\rm H}$} & \colhead{$\log U$} &
\colhead{$\log L_{\rm [O III]} = \log L_{\rm [Fe VII]}$} &
\colhead{$\log N_{\rm H}$} & \colhead{[Fe X]/[Fe VII]\tablenotemark{a}} \nl
 (cm$^{-3}$) & & (erg s$^{-1}$) &  (cm$^{-2}$) & 
}
\startdata
7.0 & $-1.4$ & 38.8 & 21.1 & 53  \nl
7.0 & $-1.6$ & 39.1 & 20.8 & 12  \nl
7.0 & $-1.8$ & 39.1 & 20.3 & 3.0 \nl
7.5 & $-1.4$ & 38.9 & 21.3 & 57  \nl
7.5 & $-1.6$ & 39.2 & 21.0 & 14  \nl
7.5 & $-1.8$ & 39.4 & 20.8 & 3.4 \nl
7.5 & $-2.0$ & 39.5 & 20.4 & 0.8 \nl
8.0 & $-1.4$ & 38.6 & 21.3 & 111 \nl
8.0 & $-1.6$ & 39.0 & 21.1 & 21  \nl
8.0 & $-1.8$ & 39.3 & 20.3 & 5.2 \nl
8.0 & $-2.0$ & 39.3 & 20.6 & 1.5 \nl
8.0 & $-2.2$ & 39.1 & 20.1 & 0.4 \nl
8.5 & $-1.4$ & 38.3 & 21.3 & 163 \nl
8.5 & $-1.6$ & 38.6 & 21.1 & 52  \nl
8.5 & $-1.8$ & 38.9 & 20.9 & 10  \nl
8.5 & $-2.0$ & 39.0 & 20.6 & 2.8 \nl
8.5 & $-2.2$ & 38.8 & 20.3 & 0.7 \nl
9.0 & $-1.6$ & 38.2 & 21.1 & 77  \nl
9.0 & $-1.8$ & 38.4 & 20.9 & 20  \nl
9.0 & $-2.0$ & 38.5 & 20.6 & 5.3 \nl
9.0 & $-2.2$ & 38.4 & 20.3 & 1.4 \nl
\enddata
\tablenotetext{a}{The intensity ratio of [FeX]$\lambda$6374/[FeVII]$\lambda$6087.}
\end{deluxetable}
\clearpage

\begin{deluxetable}{ccccccl}
\tablecaption{Results of the dual-component models}
\tablehead{
\colhead{$\frac{\displaystyle f_{\rm [O III]}({\rm Torus})}{\displaystyle f_{\rm [O III]}({\rm Total}
)}$} &
\colhead{$\log L_{\rm [O III]}$} &
\colhead{$\log L_{\rm [O I]}$} &
\colhead{$\log L_{\rm [Fe VII]}$} &
\colhead{$\log (\frac{\displaystyle {\rm [O I]}}{\displaystyle {\rm [O III]}})$} &
\colhead{$\log (\frac{\displaystyle {\rm [Fe VII]}}{\displaystyle {\rm [O III]}})$} &
\colhead{Model} \nl
 & (erg s$^{-1}$) & (erg s$^{-1}$) & (erg s$^{-1}$) &  &  &  
}
\startdata
\multicolumn{7}{c}{NLR}\nl
\hline
 & 42.3 & 41.1 & 40.1 & $-1.2$ & $-2.2$ & ~\tablenotemark{a} \nl
\hline
\multicolumn{7}{c}{Torus}\nl
\hline
 & 39.5 & \nodata & 39.5 & \nodata & 0 &
    ~\tablenotemark{b} \nl
\hline
\multicolumn{7}{c}{NLR + Torus}\nl
\hline
0.02 & 41.2 & 40.0 & 39.6 & $-1.2$ & $-1.6$ &
      ~\tablenotemark{c} \nl
0.10 & 40.5 & 39.2 & 39.5 & $-1.3$ & $-1.0$ &
      ~\tablenotemark{d} \nl
0.20 & 40.2 & 38.9 & 39.5 & $-1.3$ & $-0.7$ &
      ~\tablenotemark{e} \nl
0.50 & 39.8 & 38.3 & 39.5 & $-1.5$ & $-0.3$ &
      ~\tablenotemark{f} \nl
\enddata
\tablenotetext{a}{$\log n_{\rm H}$ (cm$^{-3}$) = 4.0 and $\log U=-2.0$.}
\tablenotetext{b}{$\log n_{\rm H}$  (cm$^{-3}$) = 7.5, $\log U=-2.0$, 
and $\log N_{\rm H}$  (cm$^{-2}$) = 20.4.}
\tablenotetext{c}{${\rm NLR}\times 0.078 + {\rm Torus}$.}
\tablenotetext{d}{${\rm NLR}\times 0.014 + {\rm Torus}$.}
\tablenotetext{e}{${\rm NLR}\times 0.0064 + {\rm Torus}$.}
\tablenotetext{f}{${\rm NLR}\times 0.0016 + {\rm Torus}$.}
\end{deluxetable}

\clearpage


\figcaption{%
 Single-cloud photoionization models are shown in this diagram
of [OI]$\lambda$6300/[OIII]$\lambda$5007 vs
[FeVII]$\lambda$6087/[OIII]$\lambda$5007.
The circles are S1s, the triangles are S1.5s, and the squares are
S2s. The filled symbols are the objects with [FeX] emission while
the open symbols are the objects without [FeX] emission.
The numbers labeling  the lowest dashed line represent $\log n_{\rm H}$.
}

\figcaption{%
New dual-component photoionization models are shown in this diagram
of  [OI]$\lambda$6300/[OIII]$\lambda$5007 vs
[FeVII]$\lambda$6087/[OIII]$\lambda$5007.
Symbols have the same meaning as in Figure 1.
The numbers labeling  the lowest dashed line represent $\log n_{\rm H}$.
}

\clearpage

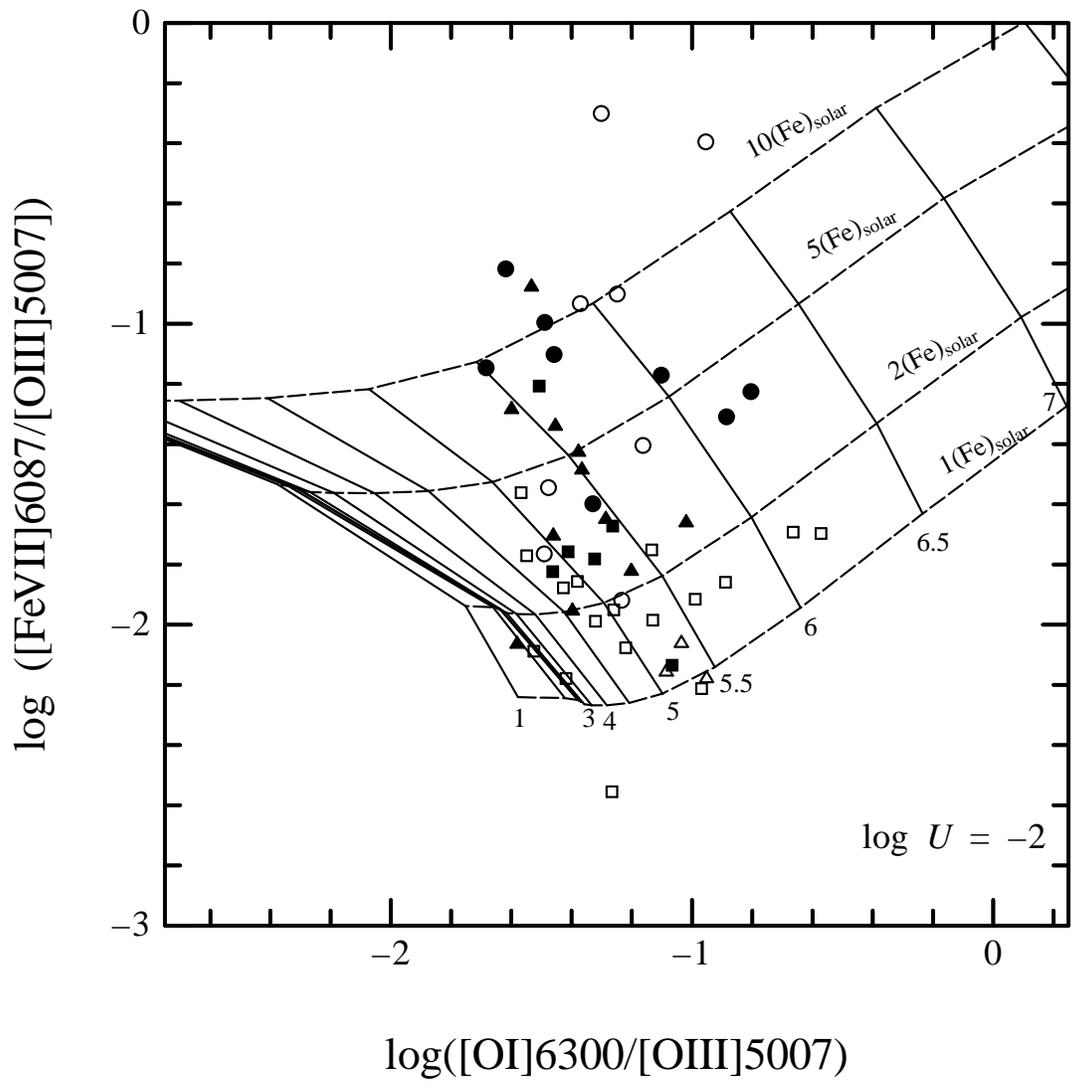
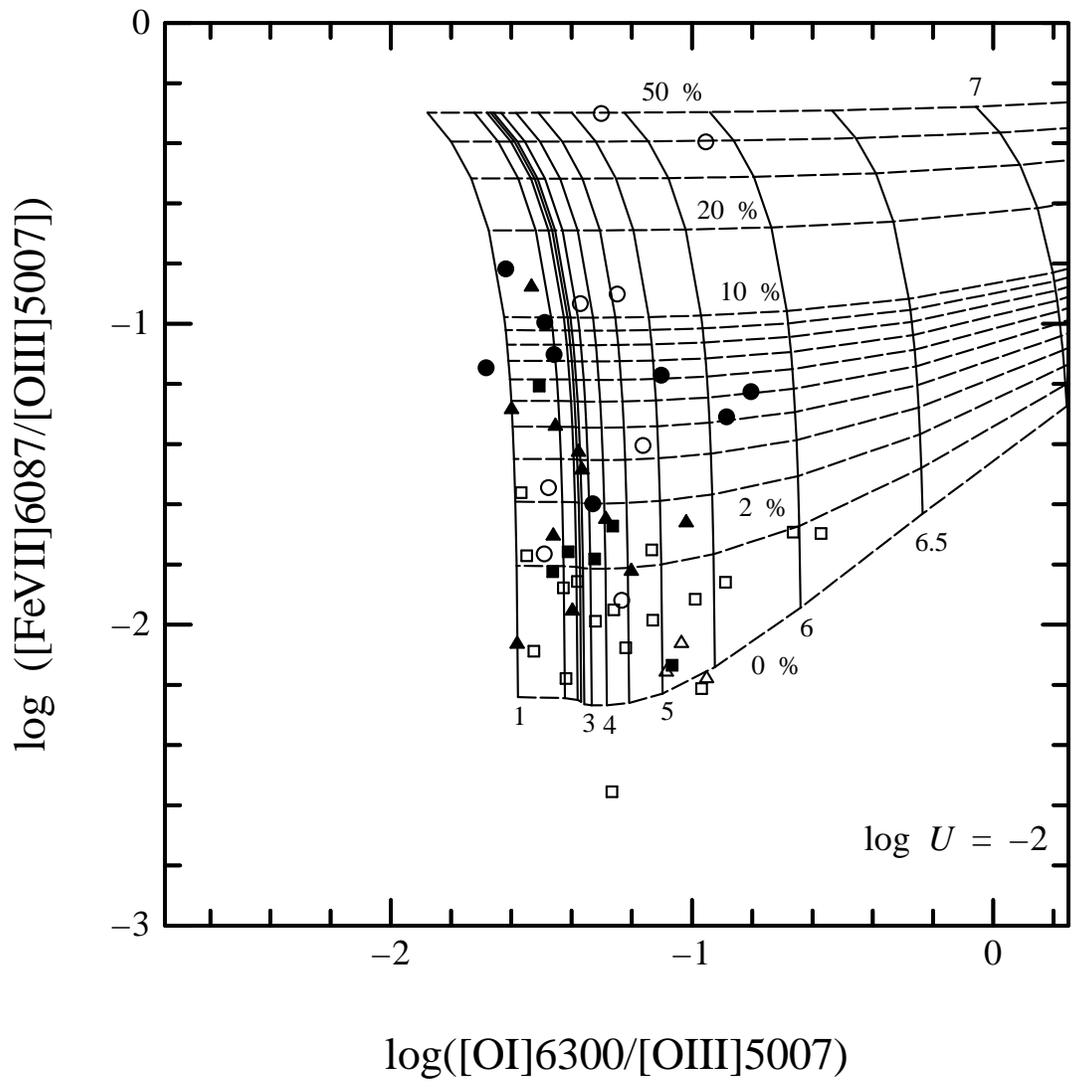
\end{document}